\documentclass[prd,aps,showpacs,epsf,floats,onecolumn]{revtex4}%
\usepackage{amssymb}
\usepackage{amsfonts}
\usepackage{amsmath}
\usepackage{graphicx}%
\providecommand{\U}[1]{\protect\rule{.1in}{.1in}}
\begin{document}
\title{\textbf{Feynman's \textquotedblleft Simulating Physics with
Computers\textquotedblright}}
\author{\textbf{Paul M. Alsing}$^{1}$, \textbf{Carlo Cafaro}$^{1,2}$, \textbf{Stefano
Mancini}$^{3,4}$}
\affiliation{$^{1}$University at Albany-SUNY, Albany, NY 12222, USA}
\affiliation{$^{2}$SUNY Polytechnic Institute, Utica, NY 13502, USA}
\affiliation{$^{3}$University of Camerino, I-62032 Camerino, Italy}
\affiliation{$^{4}$INFN,\ Sezione di Perugia, 06123 Perugia, Italy}

\begin{abstract}
This invited essay belongs to a series considering highly influential articles
published by the \emph{International Journal of Theoretical Physics}.

In this paper, we highlight the physical content and the profound consequences
of Richard Feynman's 1982 paper on \textquotedblleft\emph{Simulating Physics
with Computers}\textquotedblright.

\end{abstract}

\pacs{Quantum Computation (03.67.Lx), Quantum Information (03.67.Ac).}
\maketitle

Feynman delivered his seminal lecture \textquotedblleft Simulating physics
with computers\textquotedblright\ at the Physics of Computation Conference at
the MIT Endicott House in May of 1981. The content of his lecture was later
published by the International Journal of Theoretical Physics in 1982
\cite{feynman82}. Nowadays, Feynman's 1982 paper is properly regarded as one
of the most influential pieces of work that helped laying out the foundations
of quantum computing as a research discipline in its own right.

In Ref. \cite{feynman82}, Feynman starts from the observation that natural
phenomena are quantum rather than classical in their essence. Therefore, he
focuses on the simulation of quantum systems. He employs counting arguments to
preliminarily deduce that probabilistic simulations of quantum systems by
means of classical digital computers with local connections are not efficient
since they use resources that do not scale well with the size of the system.
He then further supports his deduction by focusing on the quantum mechanics of
a two-photon correlation experiment. This experiment can be described as
follows. Consider a source (an atom, for instance) that emits two maximally
entangled photons in opposite directions (one going to the right and the other
to the left, for instance) toward detector-$1$ and detector-$2$
\cite{feynman82,peres95}. An analyzer is placed before each detector so that a
photon polarized along the $x$-axis (horizontal polarization) goes through
detector-$1$, while a photon polarized along the $y$-axis (vertical
polarization) is blocked. Analogously, a photon polarized along the $y$-axis
goes through detector-$2$, while a photon polarized along the $x$-direction is
blocked. Each analyzer is assumed to be rotated independently of the other.
For example, when the transmission axis of analyzer-$1$ is rotated by the
angle $\phi_{1}\geq0$, analyzer-$2$ is set to $\phi_{2}\geq0$. Choosing
$\phi_{2}-\phi_{1}=30^{o}$, Feynman states that quantum mechanical predictions
agree with experimental results. Specifically, the probability of agreement
between the two detector outputs (i.e., both collect a photon or neither does)
calculated with standard quantum mechanics techniques equals $\cos^{2}\left(
\phi_{2}-\phi_{1}\right)  =3/4=75\%$. Subsequently, Feynman also shows
essentially that quantum theory is not compatible with a local hidden-variable
theory of classical type since imposing this compatibility leads to
theoretical predictions that are in disagreement with experimental
observations. Indeed, admitting a realistic local hidden variable description
of quantum mechanics, Feynman makes use of clever counting arguments to show
that the probability of agreement between the two detector outputs is
necessarily less than $2/3\approx67\%<75\%$. The value $2/3$ emerges from the
fact that when $\phi_{2}-\phi_{1}=30^{o}$, the probability of agreement is
given by the fraction of experiments in which the readings of the two
detectors agree and equals the ratio between the number of successful
experiments (i.e., $8$ in the best possible scenario) and the number of
possible experiments (i.e., $12$) \cite{man22}. Remarkably, without ever
mentioning Bell \cite{whitaker16}, Feynman offers in Ref. \cite{feynman82} an
alternative demonstration of Bell's Theorem \cite{bell64}. In a sense, this
comes as no surprise today given that Feynman's approach to quantum theory is
well-known for being simple and extremely powerful \cite{goyal14}. In view of
this discrepancy between experiments and local hidden-variable theories of
classical type, Feynman concludes that in order to efficiently simulate large
quantum systems specified by highly entangled states, one needs a computer
itself built of quantum mechanical elements that are governed by quantum
mechanical rules. With this conclusion, Feynman envisions the concept of a
quantum computer. For this reason, Ref. \cite{feynman82} can be justly
regarded as one of the most influential written pieces of work that set in
motion the hunt for a physical realization of a quantum computer
\cite{preskill23}. \bigskip

Despite the relevance of Feynman's insights, the content of the work in Ref.
\cite{feynman82} does not offer specific technical details on how a quantum
mechanical computer was presumed to run or how a simulation itself would come
to a physical realization \cite{nori14}. Of course, Feynman arrived at his
work in Ref. \cite{feynman82} after quite some deep thinking with a sound
knowledge of the existing models of computation \cite{fredkin82} together with
their most relevant thermodynamical aspects \cite{bennett82}. Furthermore,
among other things, he was rather mindful of the fact that the physical
realization of such a quantum computing machine would have been rather
challenging. In Ref. \cite{feynman59}, for instance, he points out the
significance of deepening our understanding about how to manipulate and
control the arrangements of things (for instance, spins) on an atomic scale.
Specifically, he argues that the meaningfulness of efficiently performing
these tasks permits to get access to an enormously greater range of possible
chemical and physical properties that materials and substances can exhibit. In
Ref. \cite{feynman85}, Feynman presented a new work on the relation between
physics and computation where he tried to describe with more technical details
(compared with the more abstract and general content of the work in Ref.
\cite{feynman82}) the functioning of an ideal quantum computer along with the
design of quantum algorithms to be run on it. Despite the assumed ideality of
the quantum machine analyzed, Feynman warns (yet again) his readers that this
quantum computer seems to be rather delicate, and the presence of
imperfections may cause considerable damage in practical scenarios. Indeed, in
actual scenarios, the quantum computer is expected to interact with the
external world, both for putting data in and for taking data out. Focusing on
the register for holding the data, for instance, there may be issues of
cross-talk, interactions between one atom and another in the register, or
interactions of atoms in the register directly with computer components that
are responsible for what is occurring along the program line, that one did not
exactly factor in. In other words, Feynman states that one needs to be mindful
of the fact that there may be small terms in the actual Hamiltonian besides
the ones defining the ideal Hamiltonian of the quantum-mechanical machine. To
combat some of these imperfections and noisy sources of errors causing loss of
coherence (i.e., decoherence) in quantum systems, Feynman points out that
error correcting codes developed for \textquotedblleft
normal\textquotedblright\ computers might help mitigating the number of errors
that can happen during the run of an algorithm on a quantum machine
\cite{feynman85}\textbf{.}\bigskip

Finally, Feynman also kept thinking about the meaning of negative
probabilities in science \cite{feynman87}, an issue that he mentioned in Ref.
\cite{feynman82} while discussing the incompatibility of quantum theory with a
local hidden-variable theory of classical type. Originally, this concept of
negative probability captured Feynman's attention while attempting to quantize
electrodynamics with cutoffs (a work on quantum electrodynamics that jointly
gave Feynman, Julian Schwinger, and Sin-Itiro Tomonaga the Nobel prize in
physics in 1965, seventeen years before his 1982 IJTP paper). In Ref.
\cite{feynman87}, Feynman argues that while the final probability of an
observable physical event must be positive, a final negative probability
simply means that scenario is not attainable or not experimentally verifiable.
Most interestingly, during a calculation of probabilities of physical events
or states, conditional probabilities and probabilities of imagined
intermediary events or states may be negative. Most compellingly for our
specific discussion on quantum simulations, as we shall discuss later, the
degree of negativity of (quasi) probabilities happens to play a key role in
determining the upper bounds on the experimental conditions for which
classical simulations of sufficiently noisy quantum systems can be efficiently
performed \cite{caves16}.\textbf{ }\bigskip

What happened during the last forty years? From Feynman's 1982 proposal
\cite{feynman82}, the amount of research and progress made by a number of
experts in a range of areas towards the physical realization of a fully
functioning quantum computer has been impressive. However, during these years,
there has been a persistent bouncing from excitement to dismay. We passed from
the level of \textquotedblleft conjecture\textquotedblright\ to the level of
\textquotedblleft proof\textquotedblright\ for the universality of quantum
simulators \cite{lloyd96}. Indeed, Lloyd described in Ref. \cite{lloyd96} how
to perform an operation on a quantum computer by providing an approximate
decomposition of the time evolution operator for a quantum system of
interacting particles by means of a short sequence of elementary gates. The
first two historical \textquotedblleft textbook quantum
algorithms\textquotedblright\ of practical importance in which a quantum
computer exhibited a substantial speedup over the best-known classical
approaches were invented by Shor (superpolynomial speedup in factoring large
integer numbers \cite{shor97}) and Grover (quadratic speedup in an\textbf{
}unstructured database search \cite{grover97}). Moreover, the first quantum
codes for combatting the loss of coherence caused be noisy errors in quantum
computational tasks in a quantum computer were constructed by Shor
\cite{shor95,calderbank96}, Calderbank \cite{calderbank96}, and Steane
\cite{steane96}. \bigskip

In Ref. \cite{feynman82}, Feynman mentioned the potential difficulties with
simulating a quantum system of interacting fermions. Feynman was probably
referring to the \textquotedblleft sign problem\textquotedblright\ that
happens in the numerical simulation of fermionic or frustrated models by means
of Monte Carlo simulations \cite{loh90}. The origin of the problem resides in
the fact that when one maps the quantum system to an equivalent classical
system, this mapping can give rise to configurations with negative Boltzmann
weights (ultimately, a consequence of the Pauli uncertainty principle). This
situation, in turn, results in an exponential growth of the simulation time
with the number of fermionic particles (a consequence, this, of the
exponential growth of the statistical error during the numerical simulations).
This sign problem was shown to be \textrm{NP}-hard in Ref. \cite{troyer05}.
Fortunately, a quantum version of the Metropolis algorithm was proposed in
\cite{temme11}. This quantum algorithm allows to avoid the sign problem that
occurs in classical simulations thanks to the direct sampling from the
eigenstates of the Hamiltonian of the quantum system whose equilibrium and
static properties one wishes to simulate. In Ref. \cite{feynman82}, Feynman
estimated that one needs generally an exponentially large number of steps to
simulate a quantum process on a classical computer. \bigskip

In his argumentation, Feynman also discusses the properties of the phase-space
quasiprobability distribution (i.e., the Wigner distribution). This
quasiprobability distribution is almost a classical probability distribution
since, unlike classical probabilities, it can assume negative values that
underline the quantum-mechanical nature of the physical system being
considered. Interestingly, it was reported in Ref. \cite{caves16} that for
sufficiently large loss and noise in the system (i.e., photons) where the
quantum process (i.e., boson sampling) occurs, the Wigner function can become
positive. Then, thanks to this achieved positivity, an ordinary classical
simulation of the quantum process can be performed efficiently without the
need of an exponentially large number of computational steps. The findings of
Ref. \cite{caves16} are significant since, without suppressing Feynman's 1982
suggestion about the necessity for quantum simulation, they specify upper
bounds on the experimental conditions for which an ordinary classical computer
can work efficiently (with, ideally, a runtime that scales polynomially with
the size of the physical system) as a simulator of certain classes of quantum
systems and quantum processes \cite{franson16}. Note that while the Wigner
distribution is a valid positive probability distribution on Gaussian states
(for example, the ground state of the harmonic oscillator and coherent states)
and plays a fundamental role in quantum information and metrology
\cite{wang07}, for the purposes of quantum simulation and computation,
non-Gaussian states are desired. \bigskip

So, when is a quantum-mechanical simulation envisioned in Feynman's 1982 IJTP
paper strictly necessary? When does it show a clear advantage over the best
classical simulations? There are important questions that, unfortunately, need
much more work to be addressed in a meaningful manner \cite{ball22}. Part of
the problem resides in the fact that two major challenges with quantum
computers are the identification of good a quantum computer technology and the
selection of a suitable set of scientific applications for quantum simulations
\cite{alex21}. Current technologies for the physical implementation of quantum
simulations include trapped ions, electronic spins (quantum dots),
superconducting circuits, photons (linear optics), and nuclear spins
(NMR-nuclear magnetic resonance). Each technology has its own strengths and
weaknesses, as observed from actual experimental realizations \cite{nori14}.
Two of the most important performance measures are controllability (i.e., the
ability of controlling and measuring each individual qubit) and scalability
(i.e., controlling an array of at least a few tens of qubits). The healthy
competition between these different technological platforms for quantum
simulations is not a \textquotedblleft winner takes all\textquotedblright%
\ scenario. Each platform has its own advantages and limitations, and
different approaches often address complementary features of quantum
simulation. \bigskip

By a quantum computer, one usually implies a (universal) quantum machine able
of performing full blown quantum computational tasks, including running
general algorithms, classical data on quantum devices, or quantum data on
quantum devices. The most prominent model of universal quantum computation is
the gate-based quantum computing model \cite{rieffel14}. The process of
applying these gates is commonly considered to be controlled by an algorithm
that runs on a digital computer \cite{nielsen00}. A fact that is reflected
into the emergence of quantum programming languages. Indeed, in recent years,
there has been the making of several open source quantum computing program
languages such as Google's Cirq, IBM's QisKit, Microsoft's Quantum Development
Kit, and Rigetti's Quil \cite{alex21}\textbf{.} However, tasks like compiling
quantum algorithms to gate-circuits, or compiling quantum gate-circuit
emulation functions, should be analyzed in terms of a rigorous mathematical
theory of computability \cite{el16}. This, in turn, might hide serious
difficulties and hinder the possibility for quantum information processing
hardware to become relevant in practice. Returning to the concept of quantum
simulations, we remark that it is essentially possible to identify three types
of simulation \cite{nori14}: i) digital simulation; ii) analog quantum
simulation (including, for instance, quantum annealing \cite{das08}); iii)
quantum-information inspired algorithms for the classical simulation of
quantum systems.\textbf{ }It is clear that within this classification of
types, a (universal) quantum computer can function as a quantum simulator.
However, one can also use as a quantum simulator an alternative device built
specifically for the simulation, for instance a quantum annealer. Then, a
digital simulation is intended to run on a (universal) quantum computer, while
the analog quantum simulation is performed on an application-specific device.
Finally, a quantum information inspired simulation generally runs on
(universal) a classical computer. In general, during a quantum simulation, a
quantum simulator is a controllable quantum system used to simulate other
quantum-mechanical systems (for instance: time evolution, ground state, energy
levels, thermal energy, heat capacity, correlation functions, and other static
and dynamic observable properties). A successful digital quantum simulation,
for instance, happens when the initial-state preparation, the implementation
of the time evolution, and the measurement are realized employing solely
polynomial resources \cite{nori14}. In Ref. \cite{feynman82}, Feynman requires
that for an exact simulation to happen, the quantum-mechanical computer should
do exactly as Nature. This statement is reminiscent of what today is known as
analog quantum simulation. In this type of simulation, the Hamiltonian
\textrm{H}$_{sys}$ of the system is directly mapped onto the (controllable)
Hamiltonian \textrm{H}$_{sim}$ of the simulator, \textrm{H}$_{sys}%
\leftrightarrow$\textrm{H}$_{sim}$. The third type of simulation, instead, is
characterized by the use of methods of quantum information theory into
classical numerical algorithms for the simulation of quantum many-body
systems. Interestingly, the above-mentioned sign problem insinuated by Feynman
in Ref. \cite{feynman82} was addressed in the context of this type of quantum
simulation in Ref. \cite{temme11}. \bigskip

In theory, it seems that a clear quantum advantage is expected to occur in the
simulation of the quantum dynamics (i.e., solving the time-dependent
Schr\"{o}dinger equation) of highly entangled systems of many particles,
especially chaotic quantum systems characterized by a significant temporal
rate of change of entanglement \cite{preskill18}. Indeed, when the Hamiltonian
is local, simulating the temporal evolution with a quantum computer scales
polynomially with the number $n$ of qubits in the system. The best classical
algorithms, instead, are specified by a runtime that scales exponentially with
the number of qubits $n$. As a matter of fact, the unitary evolution matrix
$U\left(  t\right)  $ is exponentially large, since its size scales like
$2^{n}\times2^{n}=e^{n\ln\left(  4\right)  }\sim e^{n}$. Specific applications
of quantum simulations are expected to occur in several fields of science,
including condensed matter, high-energy physics, cosmology, atomic physics,
nuclear physics, and quantum chemistry \cite{nori14}, as well as in the
information sciences such as machine learning \cite{biamonte17}. Hubbard
models, spin systems, quantum phase transitions, spin glasses,
superconductivity, and topological order are excellent condensed matter
phenomena for which quantum simulations are expected to be of great usefulness
\cite{nori14}. For example, an exact simulation of a one-dimensional
transverse Ising spin chain on Rigetti's and IBM's quantum processors was
performed in Ref. \cite{alba18}. Specifically, the expected value of the
transverse magnetization of the ground state of a $n=4$ Ising chain as a
function of the transverse field strength was simulated with Rigetti's
$19$-qubit processor (processor Acorn with quantum software pyQuil) and the
two IBM $5$-qubit and $16$-qubit processors (\textrm{ibmq}$\times4$ and
\textrm{ibmq}$\times5$, respectively, with software QisKit). Furthermore,
using these same three quantum processors, the time evolution simulation of
the transverse magnetization for the state $\left\vert \uparrow\uparrow
\uparrow\uparrow\right\rangle $ of a $n=4$ Ising chain was executed. In this
second set of simulations, in particular, the magnetization turns out to be
smaller than expected in intensity. However, the simulation does reproduce the
expected oscillatory pattern \cite{alba18}.\bigskip

It is worthwhile pointing out that classical simulations are one of the
possible techniques to benchmark quantum computers, even though these
simulations are exponentially demanding due to the exponential scaling of the
Hilbert space in which the quantum state is defined, as a function of the
system size. Nonetheless, people are often interested in particular states
that belong to peculiar subspaces of the full Hilbert space. Then several
techniques were introduced over the years to represent and evolve such states,
culminating with Tensor Network Methods \cite{orus14,banuls20,cirac21}. Tensor
network methods allow to approximate a quantum state by efficiently
compressing its information, introducing a controllable error, and can be used
for the simulation of quantum computers behavior. It is still unclear where
which approach exhibits an advantageous behavior. For instance, a recent
experiment on IBM's superconducting quantum processor with $127$ qubits
claiming a quantum advantage \cite{kim23} was outperformed by tensor network
methods shortly after \cite{tindall24} in terms of accuracy and precision in
simulating a kicked Ising quantum system.\bigskip

Moreover, an excellent proving ground for studying the power of quantum
simulations is high energy physics because of its very rich quantum-mechanical
content \cite{bauer23}. In particular, the theory of strong interactions
mediated by gluons (i.e., QCD-quantum chromodynamics) has several challenging
simulation problems, given the enormous dimension of the Hilbert space
together with the intense coupling strength between the interacting components
of the larger composite quantum systems. Indeed, QCD\ theory necessitates of a
larger number of qubits per site since it is specified by six flavors of
quarks- each one with four Dirac degrees of freedom and three colors
\cite{alex21}. The application of quantum simulators to cosmology is
especially useful because, in general, cosmological phenomena are not
experimentally accessible. In this context, fascinating applications include
the quantum simulation of the Unruh effect \cite{alsing05} and the quantum
simulation of the Hawking black-hole radiation \cite{nova19,sw19}. In Ref.
\cite{alsing05}, the observation of a thermal flux of particles in vacuum by
an accelerating observer was suggested to be emulated in terms of phonon
excitation of trapped ions. In Ref. \cite{nova19}, a flowing fluid of
ultracold atoms divided into \ two connected regions is considered. In one
region, the fluid moves at a supersonic speed. In the other one, it moves at
subsonic speed. Then, the black hole event horizon is simulated by the
boundary between these two regions. Furthermore, when a pair of sound waves is
produced in the vicinity of this boundary (in analogy to a pair of particles
created from vacuum of space near a black hole event horizon), one of the
waves is absorbed into the supersonic region, and the other one is radiated
away from the region (in analogy to the Hawking radiation). Quite remarkably,
recent promising applications of quantum computing and simulation have been
proposed in the field of drug discovery \cite{cao18} and finance
\cite{herman23} as well. \bigskip

Current focus in the field of quantum simulation and computation emphasizes
the search algorithms that demonstrate a\emph{ quantum advantage}\textbf{
}over comparable classical computation, vs. \emph{quantum supremacy}, where
the latter demonstrates the execution of a quantum algorithm (not necessarily
\textquotedblleft practical\textquotedblright) that is computationally
difficult or near impossible to perform on a classical computing resource. In
Ref. \cite{arute19}, Google's $53$-qubit Weber quantum processor with
programmable superconducting qubits based on the Sycamore architecture was
used for sampling the output of a pseudo-random quantum circuit acting on $53$
working qubits arranged in a two-dimensional array such that entangling
two-qubit quantum gates can be performed on neighboring qubits in the array.
Comparing with the state-of-the-art classical supercomputers, the authors
arrived at the conclusion that, despite the artificial nature of the analyzed
quantum task, a significant quantum advantage (quantum supremacy) was achieved
\cite{arute19}. However, the very same Google's $53$-qubit Weber quantum
processor was used in Ref. \cite{taz22} in simulations of properties of actual
correlated molecules in quantum chemistry and, its performance was not equally
successful. The processor was used for two types of simulations. In the first
task, the objective was to simulate the energy states of an $8$-atom cluster
of iron (\textrm{Fe}) and sulfur (\textrm{S}) found in the catalytic core of
the enzyme nitrogenase. In the second task, the aim was to study the
collective behavior (for instance, thermal energy and heat capacity) of
magnetic spins in the crystalline material alpha-ruthenium trichloride
($\alpha$-$\mathrm{RuCl}_{3}$). Despite the fact that the largest simulations
for $\alpha$-$\mathrm{RuCl}_{3}$ performed with $11$ qubits, $310$ two-qubit
gates, and $782$ single-qubit gates were not successful, simulations with a
smaller number of two-qubit gates performed much better (most likely
attributed to less overall system noise/decoherence issues associated with
smaller circuit depth, in the absence of any currently reliable quantum error
correction implementations). Specifically, simplifying the Hamiltonians into
effective low-energy spin models and exploiting significant processing of data
from exact classical simulations of similar tractable problems, the quantum
simulations with up to $100$ two-qubit gates lead to qualitatively correct
properties of the spin structure, the spectrum of the excited-state, and the
heat capacity of the materials \cite{taz22}. \bigskip

We do not claim of having discussed in this paper the state of the art of
quantum computers. Indeed, this was not our goal. However, in an attempt to
get closer to such a state of the art, we point out that there exist nowadays
hardware platforms with an order of thousands qubits. Furthermore, digital
quantum simulations with more than $100$ qubits are performed for the study of
hadron dynamics in the Schwinger model \cite{farrell24} or, alternatively, up
to $64$ qubits for equivariant quantum neural networks investigations
\cite{grossi24}. For additional reviews on quantum computing applications in
high energy physics, we refer to Refs.\cite{meglio23,banuls20B,funcke23}. In
particular, the state of the art of quantum computing examples, algorithm
methods, and hardware challenges are discussed in Ref. \cite{meglio23}. As a
curious historical side remark, we point out that Feynman's 1982 IJTP paper
was not cited much shortly after publication in the high energy physics
community. Indeed, a very rapid increase of citations took place only starting
approximately twelve years ago. For additional interesting reviews on
algorithms and methods used in quantum computing for high energy physics as
well, including the Quantum Approximate Optimization Algorithm (QAOA) and the
Variational Quantum Eigensolver (VQE), we refer to Refs. \cite{blekos24} and
\cite{tilly22}, respectively.\bigskip

In summary, Feynman's 1982 IJTP paper has paved the way towards the current
state of quantum computing which is commonly referred to as the noisy
intermediate-scale quantum (NISQ) era \cite{preskill18}. The
intermediate-scale is specified by the modest number of qubits and gate
fidelity. Within NISQ, quantum processors can contain more that thousands
qubits. However, these NISQ\ quantum processors are neither advanced enough
yet for fault-tolerance nor large enough to achieve quantum advantage.
Moreover, NISQ quantum processors cannot yet perform continuous quantum
correction to mitigate the detrimental effects of quantum decoherence that
emerge from their interaction with the external environment. These effects, as
Feynman warned in 1982, must be taken into account for physically meaningful
quantum simulations to occur. As progress is made towards these important
goals, it seems to be peacefully agreed upon that incorporating relevant
aspects of the physical problem being investigated into the design of quantum
algorithms (in analogy, to a certain extent, to the design of approximate
quantum error correcting codes \cite{cafaroPRA14}) can significantly enhance
the performance the quantum simulators and, consequently, accelerate the
progress towards (a physically meaningful) quantum advantage \cite{clinton24}%
.\bigskip

As scientists, our responsibility is to keep progressing, keep solving
wonderful mysteries, use our scientific imagination, do what we can, think
freely, welcome doubts, and accept constructive criticisms \cite{feynman55}.
Good things happen with hard work. With some luck, even in the absence of a
new physical theory to be tested, unforeseen discoveries might arrive while
experimenting with quantum-mechanical simulators of next quantum computing generations.

\bigskip

\begin{acknowledgments}
S.M. acknowledges support from Italian Ministry of Universities and Research
under \textquotedblleft PNRR MUR project PE0000023-NQSTI\textquotedblright.
The authors thank an anonymous referee for useful comments leading to an
improved version of this manuscript. Any opinions, findings and conclusions or
recommendations expressed in this material are those of the author(s) and do
not necessarily reflect the views of their home Institutions.
\end{acknowledgments}


\begin{thebibliography}{99}                                                                                               %


\bibitem {feynman82}R. P. Feynman, \emph{Simulating physics with computers},
Int. J. Theor. Phys. \textbf{21}, 467 (1982).

\bibitem {peres95}A. Peres, \emph{Quantum Theory: Concepts and Methods},
Kluwer Academic Publishers (1995).

\bibitem {man22}M. Mansuripur, \emph{Spin-}$\emph{1}$\emph{\ photons,
spin-}$\emph{1/2}$\emph{\ electrons, Bell's inequalities, and Feynman's
special perspective on quantum mechanics}, Proceedings of SPIE 12205,
Spintronics XV; 12205OB (2022).

\bibitem {whitaker16}A. Whitaker, \emph{Richard Feynman and Bell's
Theorem},\ Am. J. Phys. \textbf{84}, 493 (2016).

\bibitem {bell64}J. S. Bell, \emph{On the Einstein Podolsky Rosen paradox},
Physics \textbf{1}, 195 (1964).

\bibitem {goyal14}P. Goyal, \emph{Derivation of quantum theory from Feynman's
rules}, Phys. Rev. \textbf{A89}, 032120 (2014).

\bibitem {preskill23}J. Preskill, \emph{Quantum Computing 40 Years Later}. In
\emph{Feynman Lectures on Computation}, Chapter 7, pag.193, Tony Hey (Editor),
CRC Press (2023).

\bibitem {nori14}I. M. Georgescu, S. Ashhab, and F. Nori, \emph{Quantum
simulation}, Rev. Mod. Phys. \textbf{86}, 153 (2014).

\bibitem {fredkin82}E. Fredkin and T. Toffoli, \emph{Conservative logic}, Int.
J. Theor. Phys. \textbf{21}, 219 (1982).

\bibitem {bennett82}C. H. Bennett, \emph{The thermodynamics of computation- a
review}, Int. J. Theor. Phys. \textbf{21}, 905 (1982).

\bibitem {feynman59}R. P. Feynman, \emph{There's plenty of room at the
bottom}. Transcript of a talk given by Feynman on December 29, 1959 at the
annual meeting of the American Physical Society at Caltech. Available at http://calteches.library.caltech.edu/47/2/1960Bottom.pdf.

\bibitem {feynman85}R. P. Feynman, \emph{Quantum mechanical computers}, Found.
Phys. \textbf{16}, 507 (1986).

\bibitem {feynman87}R. P. Feynman, \emph{Negative probability}, in
\emph{Quantum Implications: Essays in Honour of David Bohm}, eds. F. David
Peat and Basil Hiley, Routledge \& Kegan Paul Ltd, London, 1987, pp. 235--248.

\bibitem {caves16}S. Rahimi-Keshari, T. C. Ralph, and C. M. Caves,
\emph{Sufficient conditions for efficient classical simulation of quantum
optics}, Phys. Rev. \textbf{X6}, 021039 (2016).

\bibitem {lloyd96}S. Lloyd, \emph{Universal quantum simulators}, Science
\textbf{273}, 1073 (1996).

\bibitem {shor97}P. W.\ Shor, \emph{Polynomial-time algorithms for prime
factorization and discrete logarithms on a quantum computer}, SIAM J. Comput.
\textbf{26}, 1484 (1997).

\bibitem {grover97}L. K. Grover, \emph{Quantum mechanics helps in searching
for a needle in a haystack}, Phys. Rev. Lett. \textbf{79}, 325 (1997).

\bibitem {shor95}P. W. Shor, \emph{Scheme for reducing decoherence in quantum
computer memory}, Phys. Rev. \textbf{A52}, R2493 (1995).

\bibitem {calderbank96}A. R. Calderbank and P. W. Shor, \emph{Good quantum
error correcting codes exist}, Phys. Rev. \textbf{A54}, 1098 (1996).

\bibitem {steane96}A. M. Steane, \emph{Multiple-particle interference and
quantum error correction}, Proc. R. Soc. Lond. \textbf{A452}, 2551 (1996).

\bibitem {loh90}E. Y. Loh Jr., J. E. Gubernatis, R. T. Scalettar, S. R. White,
D. J. Scalapino, and R. L. Sugar, \emph{Sign problem in the numerical
simulation of many-electron systems}, Phys. Rev. \textbf{B41}, 9301 (1990).

\bibitem {troyer05}M. Troyer and U.-J. Wiese, \emph{Computational complexity
and fundamental limitations to fermionic quantum Monte Carlo simulations},
Phys. Rev. Lett. \textbf{94}, 170201 (2005).

\bibitem {temme11}K. Temme, T. J. Osborne, K. G. Vollbrecht, D. Poulin, and F.
Verstrate, \emph{Quantum Metropolis sampling}, Nature \textbf{471}, 87 (2011).

\bibitem {franson16}J. Franson, \emph{Classical simulation of quantum
systems?}, Physics \textbf{9}, 66 (2016).

\bibitem {wang07}X.-B. Wang, T. Hiroshima, A. Tomita, and M. Hayashi,
\emph{Quantum information with Gaussian states}, Physics

Reports \textbf{448}, 1 (2007).

\bibitem {ball22}P. Ball, \emph{Simulations using a quantum computer show the
technology's current limits}, Physics \textbf{15}, 175 (2022).

\bibitem {alex21}Y. Alexeev et \emph{al}., \emph{Quantum computer systems for
scientific discovery}, PRX Quantum \textbf{2}, 017001 (2021).

\bibitem {rieffel14}E. G. Rieffel and W. H. Polak, \emph{Quantum Computing: A
Gentle Introduction}, MIT Press (2014).

\bibitem {nielsen00}M. A. Nielsen and I. L. Chuang, \emph{Quantum Computation
and Quantum Information}, Cambridge University

Press (2000).

\bibitem {el16}M. B. Pour-El and J. I. Richards, \emph{Computability in
Analysis and Physics}, Cambridge University Press (2016).

\bibitem {das08}A. Das and B. K. Chakrabarti, \emph{Quantum annealing and
analog quantum computation}, Rev. Mod. Phys. \textbf{80}, 1061 (2008).

\bibitem {preskill18}J. Preskill, \emph{Quantum computing in the NISQ era and
beyond}, Quantum \textbf{2}, 79 (2018).

\bibitem {biamonte17}J. Biamonte et \emph{al}., \emph{Quantum Machine
Learning}, Nature \textbf{549}, 195 (2017).

\bibitem {alba18}A. Cervera-Lierta, \emph{Exact Ising model simulation on a
quantum computer}, Quantum 2, \textbf{114} (2018).

\bibitem {orus14}R. Orus,\emph{\ A practical introduction to tensor networks:
Matrix product states and projected entangled pair states}, Annals of Physics
\textbf{349}, 117 (2014).

\bibitem {banuls20}M. C. Banuls and K. Cichy, \emph{Review on novel methods
for lattice gauge theories}, Rep. Prog. Phys. \textbf{83}, 024401 (2020).

\bibitem {cirac21}J. I. Cirac, D. Perez-Garcia, N. Schuch, and F. Verstraete,
\emph{Matrix product states and projected entangled pair states: Concepts,
symmetries, theorems}, Rev. Mod. Phys. \textbf{93}, 045003 (2021).

\bibitem {kim23}Y. Kim et \emph{al}., \emph{Evidence for the utility of
quantum computing before fault tolerance}, Nature \textbf{618}, 500 (2023).

\bibitem {tindall24}J. Tindall, M. Fishman, E. M. Stoudenmire, and D. Sels,
\emph{Efficient tensor network simulation of IBM's Eagle kicked Ising
experiment}, PRX Quantum\textbf{\ 5}, 010308 (2024).

\bibitem {bauer23}C. W. Bauer et \emph{al}., \emph{Quantum simulation for
high-energy physics}, PRX Quantum \textbf{4}, 027001 (2023).

\bibitem {alsing05}P. M. Alsing, J. P. Dowling, and G. J. Milburn, \emph{Ion
trap simulations of quantum fields in an expanding Universe}, Phys. Rev. Lett.
\textbf{94}, 220401 (2005).

\bibitem {nova19}J. R. M. de Nova, K. Golubkov, V. I. Kolobov, and J.
Steinhauer, \emph{Observation of thermal Hawking radiation and its temperature
in an analogue black hole}, Nature\textbf{\ 569}, 688 (2019).

\bibitem {sw19}S. Weinfurtner, \emph{Quantum simulation of black-hole
radiation}, Nature \textbf{569}, 634 (2019).

\bibitem {cao18}Y. Cao et \emph{al}., \emph{Potential of quantum computing for
drug discovery}, IBM Journal of Research and Development\textbf{ 62}, 1 (2018).

\bibitem {herman23}D. Herman et \emph{al}., \emph{Quantum computing for
finance}, Nature Reviews Physics \textbf{5}, 450 (2023).

\bibitem {arute19}F. Arute et \emph{al}., \emph{Quantum supremacy using a
programmable superconducting processor}, Nature \textbf{574}, 505 (2019).

\bibitem {taz22}R. N. Tazhigulov et \emph{al}., \emph{Simulating models of
challenging correlated molecules and materials on a Sycamore quantum
processor}, PRX Quantum \textbf{3}, 040318 (2022).

\bibitem {farrell24}R. C. Farrell, M. Illa, A. N. Ciavarella, and M. J.
Savage, \emph{Quantum simulations of hadron dynamics in the Schwinger model
using 112 qubits}, arXiv:quant-ph/2401.08044 (2024).

\bibitem {grossi24}C. Tuysuz, S. Y. Chang, M. Demidik, K. Jansen, S.
Vallecorsa, and M. Grossi, \emph{Symmetry breaking in geometric quantum
machine learning in the presence of noise}, arXiv:quant-ph/2401.10293 (2024).

\bibitem {meglio23}A. Di Meglio et \emph{al}., \emph{Quantum computing for
high-energy physics: State of the art and challenges. Summary of the QC4HEP
Working Group}, arXiv:quant-ph/2307.03236 (2023).

\bibitem {banuls20B}M. C. Banuls et \emph{al}., \emph{Simulating lattice gauge
theories within quantum technologies}, Eur. Phys. J. \textbf{D74}, 165 (2020).

\bibitem {funcke23}L. Funcke, T. Hartung, K. Jansen, and S. Kuhn, \emph{Review
on quantum computing for lattice field theory}, Proceedings of Science 430,
\emph{The 39th International Symposium on Lattice Field Theory} (LATTICE2022),
228 (2023).

\bibitem {blekos24}K. Blekos, D. Brand, A. Ceschini, C.-H. Chou, R.-H. Li, K.
Pandya, and A. Summer, \emph{A review on quantum approximate optimization
algorithm and its variants}, Phys. Rep. \textbf{1068}, 1 (2024).

\bibitem {tilly22}J. Tilly et \emph{al}.,\emph{\ The variational quantum
eigensolver: A review of methods and best practices}, Phys. Rep. \textbf{986},
1 (2022).

\bibitem {cafaroPRA14}C. Cafaro and P. van Loock, \emph{Approximate quantum
error correction for generalized amplitude-damping errors}, Phys. Rev.
\textbf{A89}, 022316 (2014).

\bibitem {clinton24}L. Clinton et \emph{al}., \emph{Towards near-term quantum
simulation of materials}, Nature Communications \textbf{15}, 211 (2024).

\bibitem {feynman55}R. P. Feynman, \emph{The value of science}, Engineering
and Science \textbf{19}, 13 (1955).
\end{thebibliography}
\end{document}